# Mechanical control of vibrational states in single-molecule junctions

Youngsang Kim,[1,]* Hyunwook Song,[2] Florian Strigl,[1] Hans-Fridtjof Pernau,[1] Takhee Lee,[2] and Elke Scheer[1,†]

[1]*Department of Physics, University of Konstanz, D-78457 Konstanz, Germany*

[2]*Department of Nanobio Materials and Electronics, and Department of Materials Science and Engineering, Gwangju Institute of Science and Technology, Gwangju 500-712, Korea*

*E-mail address: youngsang.kim@uni-konstanz.de

†Electronic address: elke.scheer@uni-konstanz.de

**Abstract**

We report on inelastic electron tunneling spectroscopy measurements carried out on single molecules incorporated into a mechanically controllable break-junction of Au and Pt electrodes at low temperature. Here we establish a correlation between the molecular conformation and conduction properties of a single-molecule junction. We demonstrate that the conductance through single molecules crucially depends on the contact material and configuration by virtue of their mechanical and electrical properties. Our findings prove that the charge transport *via* single molecules can be manipulated by varying both the molecular conformation (*e.g.*, *trans* or *gauche*) and the contact material.

Extensive studies on charge transport through single molecules have been performed for the implementation of molecular-scale devices, as well as with the objective of understanding how the molecules are able to carry charges in such devices [1-6]. The conductance of the same single molecule contacted with a mechanically controllable break-junction (MCBJ) or with a scanning tunneling microscope (STM) is reported to have various values because the contact geometry and the molecular conformation may vary [7,8]. To understand precisely such dependences, more sophisticated experimental studies are required, in spite of the complexity of those measurements. In such single-molecule devices, both the contact geometry and the material of electrodes (*e.g.*, gold (Au) or platinum (Pt)) can significantly influence the charge transport through the single molecule, although it is specifically anchored by functional end groups (*e.g.*, thiol (-SH)) [9-12]. Alkanedithiol is one of the most appropriate candidates to study these properties, owing to its simple and flexible structure with the σ-bonding. Imbedded into a junction it can adopt the usual *trans* conformation as well as a defect conformation. A well-known defect is given by the so-called *gauche* conformation, which is predicted to give rise to alterations of the charge tunneling [7,13-15]. To understand such manifold behavior of a single-molecule junction, inelastic electron tunneling spectroscopy (IETS) has been introduced as a powerful tool, which is very sensitive and applicable for detecting vibrational excitation in solid-state molecular devices [4-6,16,17]. Here we discuss IETS measurements for 1,6-hexanedithiol [SH-(CH$_2$)$_6$-SH, denoted as HDT] molecules when stretching both Au and Pt MCBJs used as adjustable electrodes at low-temperature. The signature of the different molecular conformations and contact geometry are demonstrated by the appearance of particular IETS signals as well as by changes in the conductance.

The HDT molecules are connected to the electrodes formed by the MCBJ technique as illustrated in Fig. 1(a). The details of the device fabrication [18] are described in the supplementary material [19]. In order to determine the preferred conductance values of the molecular junctions, we repeatedly opened and closed the junctions and recorded the conductance histograms as shown in Fig. 1(b). The inset of Fig. 1(b) presents typical conductance traces acquired during opening processes. The molecular junctions show several plateaus of conductance as well as conductance variations within a plateau. In both histograms two clear conductance maxima are observed and denoted as 'highest conductance' (HC) and 'lowest conductance' (LC), respectively, as indicated by the arrows. For the Au-HDT-Au we observe two additional intermediate conductance peaks. We assume that these peaks arise from multiple possible contact geometries of Au junctions. The appearance of several preferred conductance values of an individual alkanedithiol molecule between Au electrodes had been observed and described theoretically before [7]. We concentrate with our IETS measurements on the HC and LC regimes to analyze distinct changes and differences between the two electrode materials. The conductance values of these maxima are approximately twofold higher in the Pt-HDT-Pt junctions than for the Au-HDT-Au junctions, in agreement with previous studies [9,10]. Since a narrow 5$d$ band of Pt is located at the Fermi level ($E_F$), the local density of states (LDOS) of the $d$ band for Pt at $E_F$ is higher by one order of magnitude than that of Au which exhibits stronger $s$-orbital contribution, resulting in an enhancement of the conductance [9-12,21]. The current-voltage ($I$-$V$) characteristics (see supplementary material Fig. S2 [19]) can be well described by the single-level model [22] for symmetric metal-molecule coupling for both junction types, Au-HDT-Au and Pt-HDT-Pt. No temperature dependence of the $I$-$V$ characteristics is observed, and thus we conclude that tunneling is the conduction mechanism for these single-molecule junctions.

Representative IETS spectra are shown in Fig. 1(c) together with curves symmetrized with respect to the bias polarity obtained by the simple formula ($y = (f(x) - f(-x))/2$). The spectra seem highly symmetric, implying that the IETS signals originate from the excitation of molecular vibrations [4-6]. The IETS spectra defined as $(d^2I/dV^2)/(dI/dV)$ are presented while separating the junction from the HC (Fig. 2(a)) to LC (Fig. 2(b)). The distance scale is set to zero at the beginning of the HC plateau. After stretching the junction for about 4.5 Å, the conductance jumps to the LC regime. We continuously stretched the junction to a total elongation of 14 Å. The distance values are calibrated by the linear-fitting of experimental tunneling curves [23]. By comparison with previously studied IETS measurements and theoretical calculations, the vibrational peaks in the spectra are assigned: I: gold-sulfur stretching ($\nu$(Au-S)), II: carbon-sulfur stretching ($\nu$(C-S)), III: carbon-hydrogen rocking ($\delta_r$(CH$_2$)), IV: carbon-carbon stretching ($\nu$(C-C)), V: carbon-hydrogen wagging ($\gamma_w$(CH$_2$)), VI: carbon-hydrogen scissoring ($\delta_s$(CH$_2$)) modes, (see also supplementary material Table S1 [19]).

The shapes of IETS spectra in HC of Fig. 2(a) and LC of Fig. 2(b) vary, because the IETS depends on the molecular conformation, the atomic arrangement, and the metal-molecule coupling. Moreover these variables can influence the molecular conductance as well [7,24,25]. Firstly, we investigated potentia changes in molecular conformation for the Au-HDT-Au junctions, *e.g.*, the *gauche* and *trans* molecular conformation. The *trans* conformation is the usual one, in which the H atoms attached to neighboring C atoms are positioned opposite to each other. The carbon chain has a zig-zag shape all over the molecule. If one *gauche* defect is present, the H atoms attached to the two neighboring C atoms are rotated along the long axis of the molecule such that they enclose an angle of roughly 120 degrees, and the C chain has a kink (see Fig. 2 d). In HDT with six carbons several *gauche* defects may appear. In simulations, the appearance of only one *gauche* effect has been shown to be sufficient to adapt the molecule to a given electrode spacing [7,13]. As shall be explained below, we attribute the spectra shown in blue to the molecule being in the *gauche* conformation. The *trans* conformation has been predicted to have higher conductance than the *gauche* conformation by roughly one order of magnitude. [7,13,14,25]. The current-carrying molecular orbital (MO) is rather delocalized in the *trans* conformation, whereas the overlap of the MOs is weaker in the *gauche* conformation, thus causing the orbitals to become staggered [7,14,26]. We analyze the conductance values obtained just before the corresponding IETS measurements, as shown in Fig. 2(e). Sudden drops or increases of the conductance are taken as criterion for a conformational change from the *trans* to the *gauche* conformation or vice versa, and are used for color-coding the spectra. The average conductance of the *trans* and the *gauche* conformation according to this criterion of 30 samples is shown in the inset of Fig. 2(e). The average conductance change between both conformations is found to be a factor of three, however for individual junctions it amounts to roughly a factor of five which is in reasonable agreement with theoretical expectations [13,14]. This interpretation is strongly supported by the IETS measurements: the red-shift of $v$(C-S) as well as the enhanced intensity of both $\delta_r$(CH$_2$) and $\gamma_w$(CH$_2$). The red-shift of $v$(C-S) vibrations can be caused by the reduction of the electron density at the C-S bonds, owing to the weaker overlap of the orbitals in *gauche* conformation [7,13,27]. In the spectra, we observe a red-shift of the $v$(C-S) peak by about 4 mV, as shown in Fig 2(c). The stronger intensity in $\delta_r$(CH$_2$) and $\gamma_w$(CH$_2$) modes is also consistent with the interpretation of the *gauche* defect, because in this conformation the C-H bonds become nearly perpendicular to the metal surface (parallel to the conduction path) and are thus easier to excite by the conduction electrons [13,27,28]. Those spectra—which are consistent with the overall conductance—indicate the *gauche* conformation, and are highlighted in blue color in Figs. 2(a) and 2(b).

In the same manner, we carried out IETS measurements to clarify the differences of the molecular behavior in Pt-HDT-Pt junctions. Figures 3(a) and 3(b) are measured in HC and LC regimes in Pt-HDT-Pt junction, respectively. The vibrational modes are the same as assigned in the Au-HDT-Au junction, except for the platinum-sulfur stretching mode ($v$(Pt-S)) [28]. The vibrational mode of $v$(Pt-S), indicating that the thiol anchor group of HDT is connected robustly with the Pt atomic electrodes, is detected for the first time using the IETS technique. The *gauche* conformation is shown in red color in both the IETS spectra and the conductance trace shown in Fig. 3(c) using the same analytic method.

To investigate the change in contact geometry arising from different electrode materials, we compared the total stretching distance. Analysis of our data shows stretching distance of 15 Å and 10 Å for Au-HDT-Au and Pt-HDT-Pt junctions, respectively, and is in good agreement with previous experimental and theoretical studies [29,30]. The averaged stretching distances are demonstrated in the supplementary material Fig. S3 [19]. These results indicate that the bonding strength of Au atoms is weaker than that of Pt atoms. This interpretation is strongly supported by the excitation of metal-phonon modes. We observed enhanced IETS peaks around 18 mV (the longitudinal Au phonon mode) in Au-HDT-Au junctions as shown in Fig. 4(a). The inset of Fig. 4(a) shows the enhancement and the red-shift of the IETS peak under stretching, which indicate that the Au atoms form a chain in one dimension [31]. Here, the intensity of the longitudinal phonon mode increases by an enhancement of the electron-phonon interaction [31]. The red-shift of the longitudinal phonon mode can be interpreted as a decrease of the elastic constant of the atomic chain. However, in Pt electrodes, we have not observed the enhancement of the longitudinal Pt phonon mode which is located at 12 mV, indicating that the Pt atoms do not form long chains under stretching.

To further reveal the influence of the contact geometry, we observe the variation of both $v$(Au-S) and $v$(Pt-S) peaks as a function of junction distance as shown in Fig. 4(b). In the HC regime, the modes are stable showing the same peak positions, whereas the peak positions change abruptly when the conductance jumps to the LC regime. Such discontinuous change in the metal-S stretching mode is caused by a motion of the end-group along the metal atom sites, such as a hopping from one metal site to another, thus influencing the conductance. It was theoretically studied that the metal-S bonding energy and the conductance can change remarkably when the bonding site is altered upon stretching the junctions [8,32]. For the slight change of $v$(Au-S) in the LC regime in Fig. 4(b), we assume that the Au-Au bonds are elongated rather than Au-S bonds, forming long chains due to the weak bonding strength of Au atoms. However, for the case of Pt-HDT-Pt, we detected a pronounced variation (maximum $\Delta E \sim 12$ mV) of the energy of the $\nu$(Pt-S) mode in the LC regime. A straightforward interpretation is the following: Upon stretching, the Pt-S bond becomes first strengthened, analogous to the frequency increase when stretching a guitar string. When the stretching force exceeds the binding force, it is weakened, because at larger distances the overlap of the atomic orbitals forming the Pt-S bond is reduced [21,29]. We demonstrate a conceptual image of this situation in the inset of Fig. 4(b). Especially in the Pt-HDT-Pt junctions, the strong additional peaks around 100 mV and 180 mV, indicated by blue arrows in the fully stretched regime in Fig 3(b), are observed at a lower energy than the original vibrational modes of the uncoupled molecule. We therefore interpret them as being the extremity modes of $\delta_r(CH_2)$ and $\delta_s(CH_2)$ [27]. The presence of extremity modes indicates that the vibrations are localized at the end of the molecular backbone, not at the center. The Pt atoms are well-known to have stronger inter-atomic bonding than the Au atoms as mentioned before [11,29,30]. When the Pt-S bonds become weakened by a further stretching of the junction, the adjacent $CH_2$ vibrational modes can be staggered, resulting in the appearance of an extremity mode of the $CH_2$ group. This experimental result proves that the IETS is very sensitive to the vibrations localized at the ends of the molecule spanned between the metallic electrodes [27].

In conclusion, we observed that conductance variations of HDT single-molecule break-junctions at low temperature are dominated by changes of the contact geometry rather than by the molecular conformation. The signature of *trans* and *gauche* conformation reveals itself in the IETS spectra, though. The choice of the electrode material is not only important for providing electronic coupling, but also plays an important role for the resulting intra-molecular strain of the molecules under stretching. The stiffness of the electrode material determines the amount by which the vibrational modes can be tuned. This study will pave the way to establishing a detailed correlation between the precise atomic arrangement and the conductance properties.

We thank Thomas Frederiksen, Wulf Wulfhekel, Jan van Ruitenbeek and Artur Erbe for fruitful discussion. This work was supported by the DFG through SFB767 and the Krupp Foundation. H.S. and T.L. thank the National Research Laboratory program from the Korean Ministry of Education, Science and Technology.

**Figure Legends**

FIG. 1. (a) Schematic illustration of MCBJ system (bottom) and a scanning electron microscope image of Au break-junction electrodes with a conceptual image of the Au-HDT-Au junction. (b) Recorded histograms of Au (black) and Pt (red) junctions, repeated 2000 and 300 times, respectively. Inset shows some representative conductance traces. (c) IETS (black) of HDT single-molecule connected with Au and Pt. For negative polarity the sign of $d^2I/dV^2$ has been inverted for better illustrating the symmetry. The red curves are symmetrized with respect to the bias polarity obtained by the simple formula ( $y = (f(x) - f(-x))/2$ ).

FIG. 2. (a), (b) The normalized IETS were measured in the HC (a) and the LC (b) regimes and displaced vertically for a better visual understanding, in the order of the junction distances in the Au-HDT-Au junction. The IETS are measured from a distance of 0.5 Å (top) to 4.5 Å (bottom) for the HC regime (a) and between 5 Å (top) to 14 Å (bottom) for the LC regime (b). The distance axis is set to zero when the single molecule contact was established (this is signaled by the moment when the conductance first dropped below 0.1 $G_0$). (c) Enlarged C-S stretching mode reported in (b). (d) The schematic diagrams of the atomic structure for *gauche* and *trans* conformation. (e) The conductance was recorded, just before the measurement of the IETS, while increasing the electrode distances. The inset shows the averages (over all junctions) with standard deviations of the conductance values on the *trans* (*T*, black) and *gauche* (*G*, blue) conformations in HC and LC of Au-HDT-Au junctions. The overlap of the error bars in *gauche* and *trans* conformation is due to the average of many data which have slightly different conductance values.

FIG. 3. (a), (b) The normalized IETS were recorded in the HC (a) and the LC (b) regimes as a function of junction distances in the Pt-HDT-Pt junction. The IETS are measured approximately between 0.2 Å (top) to 5 Å (bottom) for the HC regime (a) and between 6 Å (top) to 10 Å (bottom) for the LC regime (b). (c) Conductance values just before the measurement of the IETS, while increasing the electrode distances. The inset shows the averages over all junctions with standard deviations of the conductance values on the *trans* (*T*, black) and *gauche* (*G*, blue) conformations in HC and LC of Pt-HDT-Pt junctions.

FIG. 4. (a) IETS intensity of Au (black) and Pt (red) longitudinal phonon mode while stretching. Inset shows the enhancement and red-shift of Au phonon spectra. (b) The frequency shift of each peak position for both Au-S (black) and Pt-S (red) stretching modes as a function of separating distance. The dashed vertical line classifies the HC and LC regimes.

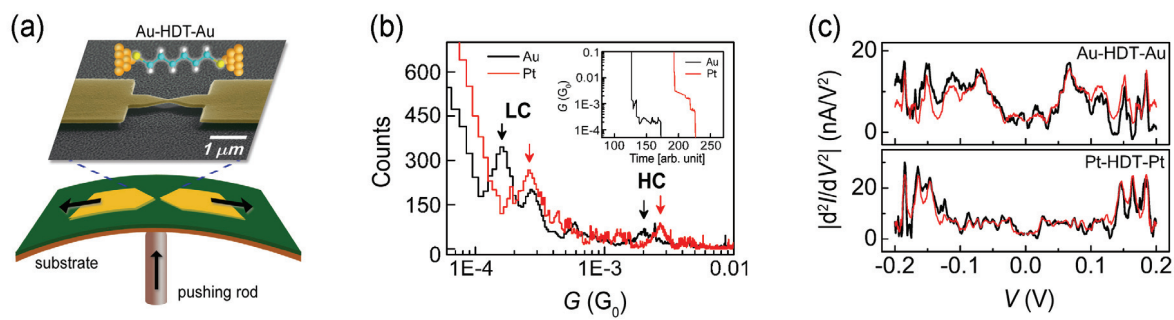

Figure 1 (Kim *et al.*)

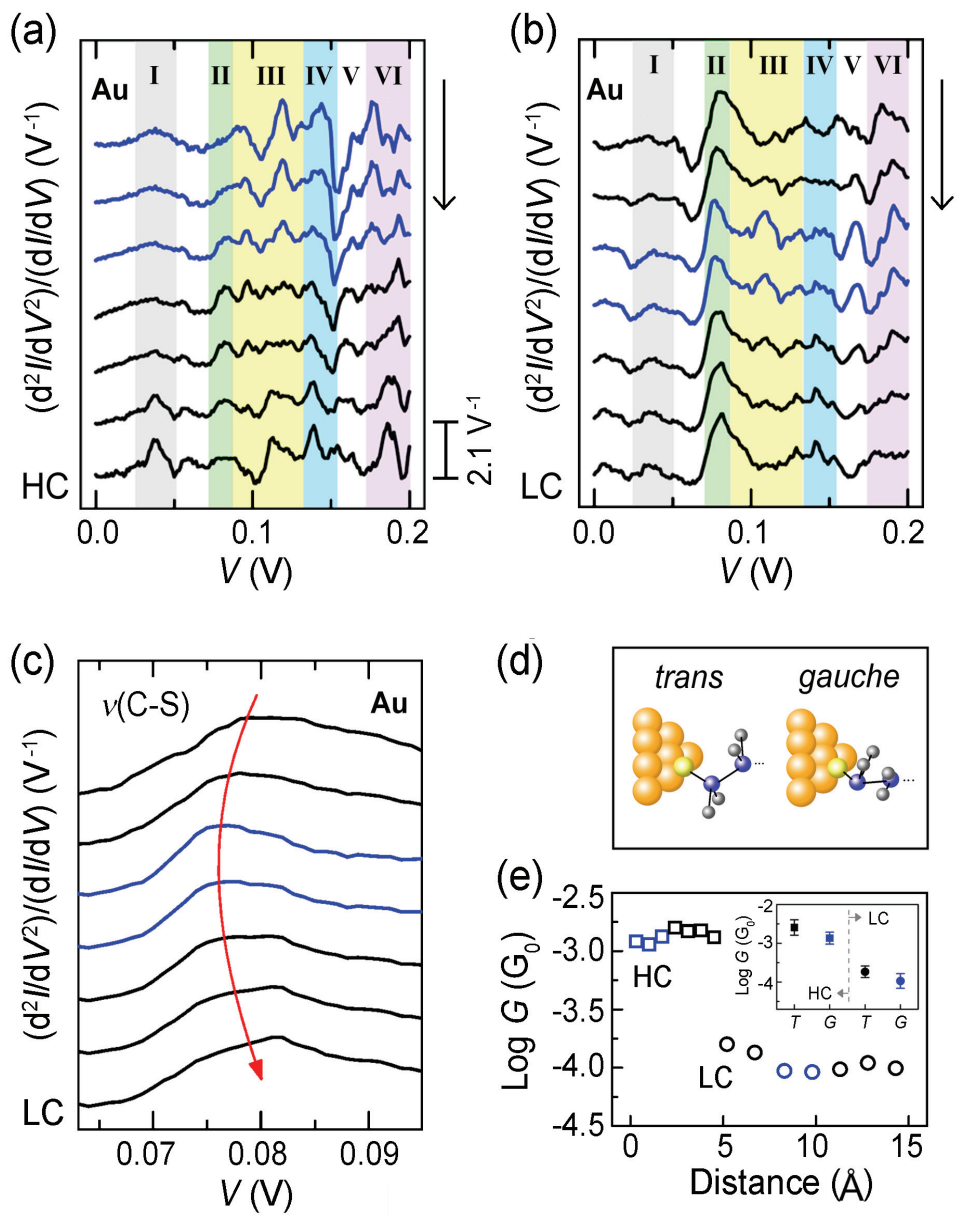

Figure 2 (Kim *et al.*)

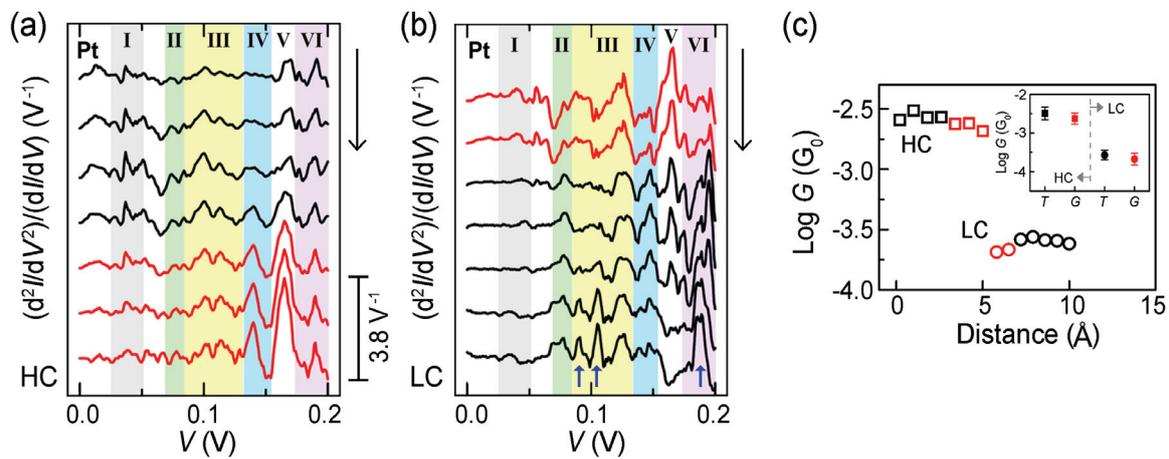

Figure 3 (Kim *et al.*)

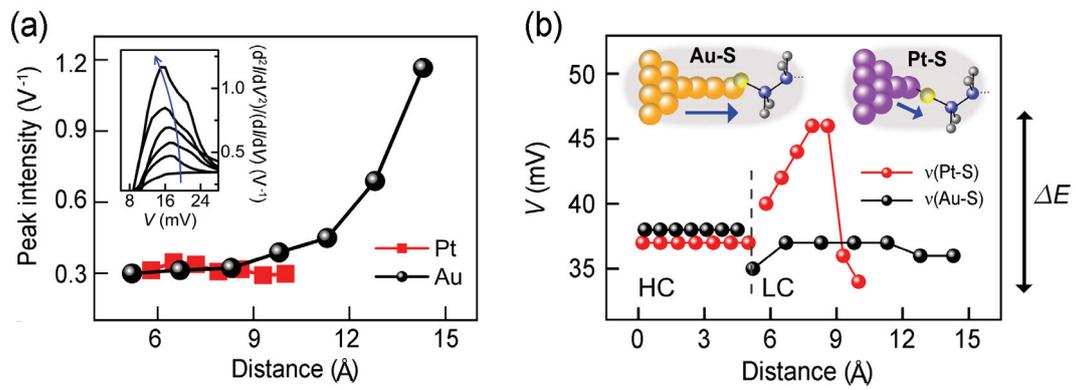

Figure 4 (Kim *et al.*)